%% file: MNechaeva.tex
\documentclass{evn2004}
\usepackage{txfonts}
\usepackage{graphicx}
\begin{document}
\include{page}
\title{
VLBI-experiments on research of solar wind plasma
}

\author{
M.\,B.\,Nechaeva\inst{1} \and V.\,G.\,Gavrilenko\inst{2} \and
Yu.\,N.\,Gorshenkov\inst{3} \and B.\,N.\,Lipatov\inst{1}
\and Liu~Xiang\inst{4} \and I.\,E.\,Molotov\inst{5}
\and A.\,B.\,Pushkarev\inst{5} \and
R.\,Shanks\inst{6}\and G.\,Tuccari\inst{7}
}

\institute{
Radiophysical Research Institute (RRI), B. Pecherskaya 25/14, 603950
Nizhnij Novgorod, Russia, e-mail: nech@nirfi.sci-nnov.ru
\and
Nizhnij Novgorod State University (UNN), Gagarina Av.,23a,603950,
Nizhnij Novgorod, Russia
\and
Special Research Bureau, MPEI, Krasnokazarmennaya 14, 111250
Moscow, Russia
\and
Urumqi Astronomical Observatory, NAO CAS, S. Beijing Road 40,
830011 Urumqi, China
\and
Central Astronomical Observatory, Pulkovo, Pulkovskoe sh. 65/1,
196140 St.-Petersburg, Russia
\and
Dominion Radio Astrophysical Observatory, P.O. Box 248, V2A 6K3
Penticton B.C., Canada
\and
Istituto di Radioastronomia, Contrada Renna Bassa, 96017 Noto,
Italy
}

\abstract{
This work  devotes to  investigations of  solar corona  and solar wind
plasma  by  the  method  of  radio  probing  with  using  of very long
baseline  interferometry   (VLBI).  We   performed  the    theoretical
calculation of  power spectrum  of interferometric  response to  radio
source  emission,  passed through  the  turbulent   medium.  Data   of
theoretical analysis are compared  with results of international  VLBI
experiments on investigations of  solar wind plasma. The  observations
were  realized  in  1998,  1999,  2000  with  participation  of  radio
telescopes, included at Low Frequency VLBI Network (LFVN): Bear  Lakes
(RT-64,  Russia),  Puschino  (RT-22,  Russia),  Urumqi (RT-25, China),
Noto   (RT-32,   Italy),   Shanghai   (RT-25,   China)   and   others.
Preprocessing  was  carried  out  with  using  of  S2  correlator   at
Penticton  (Canada).   Post  processing   of  experimental   data  was
performed at RRI (Russia) and was aimed to obtain value of solar wind  velocity
and index of spatial spectrum of electron density fluctuations.
   }

   \maketitle
%
%________________________________________________________________

\section{Introduction}

One of the most important problem of solar-terrestrial physics is  the
investigation  of  coronal  plasma~---  the  determination
of physical   parameters   of   electron
density irregularities  and  their  spatial-temporal  structure. The
method of radio sounding is widely  applied for investigation of
coronal plasma and solar wind. In this method  the signal from the
radio source  goes through  the  turbulent  medium  and  then is
received at ground-based radiotelescope.  Heterogeneous  mediums
cause  phase,  amplitude  and frequency  fluctuations  of  the
received  emission  and  distort the output signal of  the
instrument. The  analysis of these  disturbances allows  one  to  get
information  on  physical characteristics of the propagation medium.

Perspective technique in this field proves to be interferometry
method [Altunin et all, 2000, Girin et all, 1999, Spangler~S.\,R. et
all, 2002], when sounding signals are propagated through circumsolar
plasma by different paths and received at separated antennas of
VLBI-complexes. A very long baseline interferometer permits studying
the electron-density irregularities using relative phase, amplitude,
and frequency fluctuations originating on propagation paths from the
source to the receiving antennas of the interferometer. Moreover the
size of the baseline projection of the instrument to the wave
front determines the scale of irregularities (100--10000\,km),
interferometer being sensitive to them. Using of interferometric
complexes with various orientation of baseline projections allows to
obtain information about spatial structure of solar wind plasma
parameters. The observations of sources, located at different
positional angles and distances from the Sun, let us make certain
conclusions about the spatial structure and anisotropy of electron
density of investigated medium. The radiointerferometric method
allows for obtaining data by receiving both monochromatic signals
from spacecraft and wideband emission from natural radio sources.

In recent years, a number of VLBI-experiments connected with the
investigation of solar wind plasma by the method of radio sounding
has been carried out at Low Frequency VLBI Network (LFVN).

We performed theoretical analysis of power spectrum of interferometric
response to radio source emission, passed through the turbulent medium
[Gavrilenko et all, 2002, Alimov
et all, 2004].
The conclusions were compared with experimental data having for an
object to get information about physical characteristics of
propagation medium, namely, solar wind velocity and spectral index.

\section{Frequency spectrum of the interferometer response to
wideband emission under conditions of an inhomogeneous medium}

The purpose of current work is to obtain theoretical formulas describing the
response of an interferometer to the radio emission from the source,
which propagated through a turbulent medium, to simulate VLBI
experiments on radio raying of the solar wind, and to compare the
calculation results with the experimental data.

We analyzed the situation, when the noise emission of a cosmic radio
source passes along axis $z$ through the turbulent medium of solar
wind with chaotic large scale irregularities of electron density. Then
it is received by a two-element ground radio interferometer on
plane $(x,y)$, which is perpendicular to the direction of propagation.

The signals received at separated antennas having been transformed in
the tracts of the interferometer, undergo mutual correlation
processing. The power spectrum of signals at the output of the
interferometer's correlator, having information about the propagation
medium, can be represented in following way:
\begin{equation}\label{n1}
Y(\Omega_0={1\over2\pi}\int\limits_{-\infty}^{\infty}
R(\tau)\,e^{-i\Omega_0\tau}\,d\tau
\end{equation}
where
\begin{equation}\label{n2}
R(\tau)=\left\langle E_1(\vec r,t)\,E_2^*(\vec r+\vec\rho,t)\,
E_1(\vec r,t+\tau)\,E_2^*(\vec r+\vec\rho,t+\tau)\right\rangle
\end{equation}
--- the correlation function of the interferometer's output signal,
$E_{1,2}\approx e^{\Phi_{1,2}}$~--- the signals at the outputs of
the receiving tracts of the  interferometer, $\Phi$~--- the
complex function of amplitude and phase fluctuations, caused by
the turbulence, $\vec\rho$~--- the baseline of the interferometer,
$\Omega_0$~--- the frequency of Fourier-analysis.

Expression~(\ref{n2}) is described the correlation procedure, common
for interferometer reception. It has some advantages in comparison
with traditional one point receiving as it allows to investigate the
field fluctuations caused by turbulent medium only on two different
propagation paths. In this case the influence of source
self-radiation is excluded and it allows us to sound the medium not
only by monochromatic but also wideband source signal. It proves to
be very important when the later is raying by sound emission of
natural sources.

In this work the theoretical calculation of the power spectrum of the
interferometer's output signal is performed. The distribution of
field fluctuations was carried out with geometric-optical approach.
It was supposed, that the temporal changes of the electromagnetic
field parameters in the solar corona concern the
``frozen-in" hypothesis ~--- when the irregularities move from the Sun in
radial direction at the velocity of solar wind  $\vec V$. The
spatial spectrum of fluctuations of turbulence parameters
$F(\kappa)$ was described by a power law function at the wave
number range $[\kappa_0,\kappa_m]$, $\kappa_0=2\pi/\Lambda$,
$\kappa_m=2\pi/l_m$ ($\Lambda$ and $l_m$~--- are outer and inner
scale of turbulence respectively):
\begin{equation}\label{n3}
F(\kappa)=0.033\,C_N^2{(\kappa_0^2+\kappa^2)}^{-{p\over2}}\,
e^{-{\kappa_\bot^2\over\kappa_m^2}},
\end{equation}
$C_N^2$~--- the structural coefficient, determining the intensity
of fluctuations.

The spectral index is usually
considered to be equal to $p=11/3$ (Kolmogorov's spectrum) at
distances from the Sun more than $15\cdot R_\odot$ ($R_\odot$~---
radius of the Sun) and it is decreased to $p=3$ at shorter
distances. Index $p$ is the characteristics of the turbulent medium
being investigated, which is determined from the spectrum of the
interferometer's output signal.

The expression for correlation function, considering the amplitude
and phase fluctuations, has a form:
\begin{eqnarray}\label{n4}
R(\tau)&=& \exp\left\{-4\pi A^2Z\int\limits_{-\infty}^\infty
F(\vec\kappa_{\bot}) [1-\cos\vec\kappa_{\bot}\vec
V_\bot\tau]\,\times\right.\nonumber\\
&&\times\, [1-\cos(\vec\kappa_\bot\vec\rho_\bot)]d\vec\kappa_\bot+
i2\pi\,{cA^2Z^2\over\omega_0}\,\times\nonumber\\
&&\times\left.
 \int\limits_{-\infty}^\infty
\int\limits_{-\infty}^\infty\vec\kappa_\bot^2\,F(\vec\kappa_\bot)
\sin(\vec\kappa_\bot\vec
V_\bot\tau)\sin(\vec\kappa_\bot\vec\rho_\bot)\,
d\vec\kappa_\bot\right\}\!,
\end{eqnarray}
$\displaystyle A=-{\omega_p^2\over2N\omega_0c}$, $\omega_p$~---
plasma frequency, $c$~--- velocity of light, $\vec\kappa_\bot$,
$\vec\rho_\bot$, $\vec V_\bot$~--- projections of wavenumber,
baseline and solar wind velociy on the plane of the sky,
$\omega_0$~--- the central frequency of the reception, $N$~--- electron
density, $Z$~--- the thickness of irregularities' layer.

We examine two limit cases: strong and weak phase
fluctuations. In case of strong phase fluctuations, the expression
for $Y(\Omega_0)$ can be written as follows:
\begin{equation}\label{n5}
Y(\Omega_0)={1\over\sqrt{2\pi\,\langle\omega^2\rangle}}\,
\exp\left\{-{{(\Omega_0-\Delta\omega)}^2\over2\langle\omega^2\rangle}
\right\},
\end{equation}
\begin{equation}\label{n6}
\langle\omega^2\rangle=8\pi^2A^2Z\int\limits_0^\infty
\kappa^3\,F(\kappa)\,S(\kappa)\,d\kappa,
\end{equation}
\begin{equation}\label{n7}
S(\kappa)=V_x^2\left[1-J_0(\kappa\rho_x)+(V_y^2-V_x^2)\left(
{1\over2}-{1\over\kappa\rho_x}\,J_1(\kappa\rho_x)\right)\right],
\end{equation}
\begin{equation}\label{n8}
\Delta\omega={4\pi^2\over k}\,A^2Z^2\int\limits_0^\infty
\kappa^4\,F(\kappa)\,J_1(\kappa\rho_x)\,d\kappa.
\end{equation}

In (\ref{n5})--(\ref{n8}) the cross baseline projection is directed
along the axis~$x$; $J_0$, $J_1$~--- Bessel function. It is evident,
that in case of strong phase fluctuations,
the power spectrum
of the output signal $Y(\Omega_0)$ has the form of Gaussian function
independently of the choice of the spatial spectrum's type $F(\kappa)$.
Value $\langle\omega^2\rangle$ described the intensity of phase fluctuations.

The analysis confirms, that solar wind observations proved to be
more informative at weak phase fluctuations. In this case the form the
power spectrum must be drop-down power law function. When stream
velocity is directed along baseline ($\rho_y$, $V_y=0$)
dependence of spectrum from baseline $\rho_x$ and solar wind velocity
is described approximately by the expression:
\begin{equation}\label{n9}
Y(\Omega_0)=K\,{Z\over V_x}\,A^2\left[1-\cos\left(
{\Omega_0\rho_x\over V_x}\right)\right]\!\left[
\left({\Omega_0\over V_x}\right)^2+\kappa_0^2\right]^{(-p+1)/2}
\end{equation}
($K$~--- constant coefficient, (\ref{n9})~--- is received under the
condition: $\Omega_0\ll2\pi V_x \kappa_m$).

The function describing the spectrum is represented in fig.\,1 in
logarithmic scale; the dimensionless value $\Omega=\Omega_0/\Omega_v$
is marked along the abscissa axis ($\Omega_v=2\pi V_x/\rho_x$).

   \begin{figure}
  \centering
   \includegraphics[width=8cm]{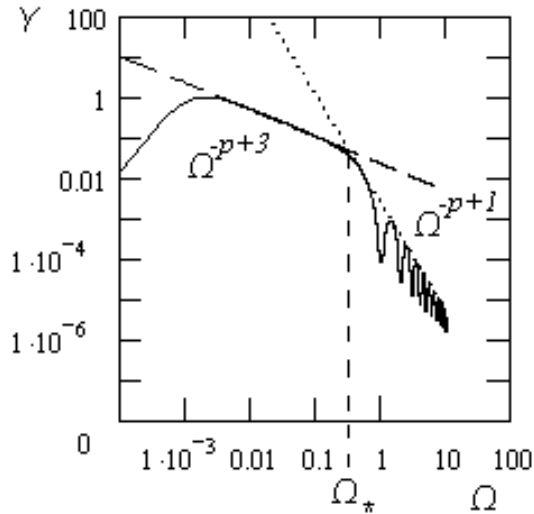}
      \caption{The power  spectrum at weak phase fluctuations
      }
         \label{Fig1}
   \end{figure}

It follows from~(\ref{n9}) that the spectrum should have oscillations
determined by the factor $\displaystyle
\left[1-\cos\left({\Omega_0\rho_x\over V_x}\right)\right]$ and
dependent only on the relationship between the solar-wind velocity
and the interferometer baseline. This makes it possible to determine
the velocity of irregularity transport on the sounding path.

Two characteristic intervals can be distinguished for this curve.
The corresponding frequency dependences are described by the next
approximation:
\begin{equation}\label{n10}
Y(\Omega)\approx{[\Omega]}^{-p+3},\quad
0<\Omega_0<\pi V_x/\rho_x,
\end{equation}
\begin{equation}\label{n11}
Y(\Omega)\approx{[\Omega]}^{-p+1},\quad
\Omega_0>\pi V_x/\rho_x.
\end{equation}

Therefore, the index $p$ of the spatial spectrum of fluctuations of
medium parameters can be determined from the slopes of the curve.
Approximating curves~(\ref{n10}) and~(\ref{n11}) intersect at some
point $\Omega_*$ from which one can determine the solar-wind velocity
if $\rho_x$ is known.

If the interferometer baseline is perpendicular to the drift
velocity of irregularities of the medium $(\rho_y\gg\rho_x)$,
then (\ref{n9}) yields
\begin{equation}\label{n12}
Y(\Omega)\approx{\rm const},\quad 0<\Omega_0<\pi V_x/\rho_x,,
\end{equation}
\begin{equation}\label{n13}
Y(\Omega)\approx{[\Omega]}^{-p+1},\quad
\Omega_0>\pi V_x/\rho_x.
\end{equation}

In this case, as well, the spectral index $p$ can also be determined
from the asymptotic dependence of the measured frequency spectrum,
while the solar wind velocity can be determined from the
characteristic frequency at the break point of measured
spectrum $\Omega_v=\pi V_x/2\rho_x$.

\section{Results of experiments}

We carried out the analysis of international VLBI experiments on
investigations of solar wind plasma, implemented in 1998, 1999, 2000.
These observations were realized with participation of radio
telescopes, included at Low Frequency VLBI Network: Bear Lakes
(RT-64, Russia), Puschino (RT-22, Russia), Hartebeesthock (RT-25,
South Africa), Arecibo (RT-305, USA), Urumqi (RT-25, China), Noto
(RT-32, Italy), Shanghai (RT-25, China), GMRT (RT-45, India). At the
course of these experiments VLBI complex received the extragalactic
source radio emission (wavelength 18 cm) passed through the turbulent
plasma of solar wind. Radio sources were located at different angular
distances from the Sun (4-37 degrees). The observational information
were recorded using the S2 system. Preprocessing of VLBI-data was
carried out at S2-correlator at Penticton (Canada). The data were
calibrated in the NRAO AIPS package using the standard technique.

   \begin{figure}
  \centering
   \includegraphics[width=8cm]{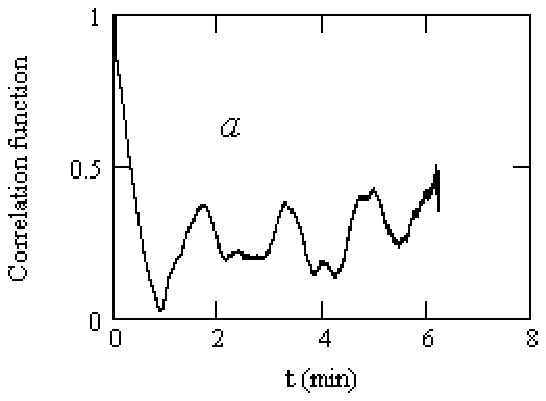}
   \includegraphics[width=8cm]{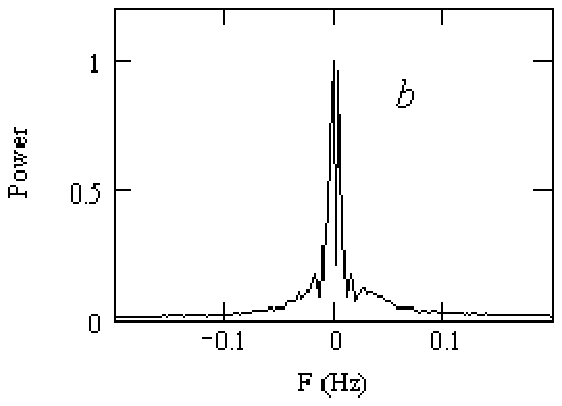}
      \caption{
The examples of the autocorrelation function (a) and power
spectrum (b) taken in the course of experiments INTAS99.4
(November,1999, source NRAO530, solar elongation $\theta=16^\circ$,
baseline Bear  Lakes (RT-64, Russia)~--- Noto (RT-32, Italy),
$\lambda=18$~cm, integration  time 0.1 sec.)
              }
         \label{Fig2}
   \end{figure}

Postprocessing was performed at RRI (Russia) and was aimed to obtain value of
solar wind velocity and index of spatial spectrum of electron density
fluctuations. The autocorrelation function of output signal was
counted. After that Fourie-analysis was performed; as a result we
have the power spectrum of field of interferometer signal. The
examples of the autocorrelation function and spectrum taken in the
course of experiments are shown at fig.~2.

The results of spectral analysis of realization series were compared
with theoretical conclusions. Obtained experimental power spectra let
us to get estimations of index of spatial spectrum $p$ according slope
of the spectrums. During observations $p$ took on a value 3.2--4.5
depending on the angular distances of radio sources to the Sun.

The determination of solar wind velocity were embarrassed by the
several reasons. The most part of experimental data was correlated
with too large integration time $(t=2$~s). This circumstance did not allow to
get output spectrum in sufficient wide band, where desired feature of
spectrum is located. Reiteration of preprocessing with integration time
$t=0.1$~s, before activity
correlator was stopped, has been implemented for sources, observed
at experiment INTAS99.4 in 1999, November.

Strong response was obtained only for one source NRAO530, located at
angular distance from the Sun $\theta=16$ degrees. Positions of other
sources, coming for solar wind observations, were too close to Sun
and their signal-to-noise relation was too low, that does not allow to
obtain distinct spectrums. The sample of power spectrum for source NRAO530
is displayed on fig.3 (baseline Bear Lakes (RT-64, Russia)~--- Noto
(RT-32, Italy), wavelength 18cm, integration time $t=0.1$~sec, 4
December, 1999). The baseline projection was directed mainly along
velocity of irregularities transfering. The conditions of it's
observations corresponded to the case of weak phase fluctuations.
Spectral index $p$, evaluated on the slope of power spectrum, is equal
to 3.8.

   \begin{figure}
  \centering
   \includegraphics[width=8cm]{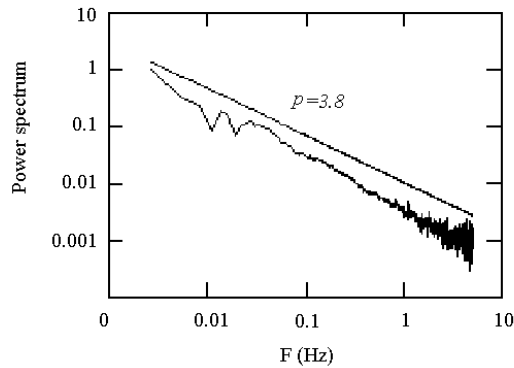}
      \caption{
 The power spectrum, obtained for source NRAO530 (experiment
INTAS99.4, 1999, $\theta=16^\circ$, baseline Bear Lakes (RT-64, Russia)~--- Noto
(RT-32, Italy), $\lambda=18$~cm, integration time 0.1 sec.); spectral index,
evaluated  from spectrum slop $p=3.8$.
              }
         \label{Fig3}
   \end{figure}

As stated above, proposed method theoretically may be used for
calculation of solar wind velocity. The expression~(\ref{n9}) shows,
that must be oscillations on spectrum wings. They depend on relation
of velocity and baseline. Experimental spectrum does not demonstrate
distinct oscillations. Supposed, velocity fluctuations at solar wind
were significant, and provoked the smoothing of minimums and
rendered impossible to evaluation of velocity. The break point of
measured spectrum could not be clear distinguished at this graph.
Further theoretical and experimental testing is need for adjusting of
procedure for solar wind velocity evaluation.

\section{Conclusions}

We performed the theoretical analysis of interferometer response on
radio emission propagated through the coronal plasma. The power
spectrum of field fluctuations was obtained using the approach of
``frozen-in" hypothesis for two cases: strong and weak phase
fluctuation. It was showed that data on the propagation medium can be
received at sounding of media by monochromatic signals from
spacecrafts, as well as wideband emission from natural radio sources.

The theoretical analysis has demonstrated that interferometer's
signal carries information about spatial spectrum of electron density
distribution, intensity of phase fluctuations and solar wind velocity.

Experimental works are satisfactory corresponded with conclusions of
theoretical analysis as regards to determination of spectral index.
Nevertheless determination of average velocity of irregularities
transportation is embarrassed by weak source signal and existing
sufficient fluctuations of velocity. With this object the program of
solar wind observations was included at schedules of last experiments
VLBR04.1, VLBR04.2, VLBR04.3, performed at Low Frequency VLBI Network
in June, July and October of 2004 at wavelength 6 cm. Records of
experimental data was
realized at system MK2. Processing of date will be made at correlator
MK2 ``NIRFI-3" at RRI, Nizhnij Novgorod, Russia. The high time
resolution of this correlator and large diameter of enabled antennas
let us to expect satisfactory results
as concerned the solar wind velocity determination.

{\em Acknowledgements}. This work is supported by grants
RFBR-02-02-39023, RFBR-04-02-27022, NSCF-10173015. We thank the staff
at the participating observatories, who made these observations
possible.

\end{document}

%% file: page.tex
\setcounter{page}{333}